\documentclass[twocolumn,prb,showpacs,fleqn,superscriptaddress]{revtex4}
\usepackage{epsfig}
\usepackage{graphicx}
\usepackage{amsfonts}
\usepackage{bm}
\usepackage{color}

\newcommand{\ct}{\cite}
\newcommand{\bi}{\bibitem}

\newcommand{\non}{\nonumber}

\newcommand{\be}{\begin{equation}}
\newcommand{\ee}{\end{equation}}
\newcommand{\ba}{\begin{eqnarray}}
\newcommand{\ea}{\end{eqnarray}}

\begin{document}

\title{Fidelity susceptibility and general quench near an anisotropic quantum critical point}

\author{Victor Mukherjee}
\email{victor@iitk.ac.in}
\author{Amit Dutta}
\email{dutta@iitk.ac.in}
\affiliation{Department of Physics, Indian Institute of Technology Kanpur, Kanpur 208 016, India}

\begin{abstract}
We study the scaling behavior of  fidelity susceptibility density $(\chi_{\rm f})$ at or close to
 an anisotropic quantum critical point characterized by two different correlation length exponents $\nu_{||}$ and $\nu_{\bot}$ along parallel and perpendicular 
spatial directions, respectively. Our studies  show that the response of the system due to a small change in the Hamiltonian near an anisotropic 
quantum critical point is different from that seen near an isotropic quantum  critical point. In particular, for a finite system with linear dimension $L_{||}$ ($L_{\bot}$) 
in the parallel (perpendicular) directions, the maximum value of $\chi_{\rm f}$  is found to increases in a  power-law fashion with $L_{||}$ for small $L_{||}$,  with an exponent
 depending on both $\nu_{||}$ and  $\nu_{\bot}$ and eventually crosses  over to a scaling with $L_{\bot}$ for $L_{||}^{1/\nu_{||}} \gtrsim L_{\bot}^{1/\nu_{\bot}}$. We also propose
scaling relations of heat density and defect density generated following a quench starting from an anisotropic quantum critical point and connect them to a generalized
fidelity susceptibility. These predictions are verified exactly both analytically and numerically taking the example of a Hamiltonian showing a semi-Dirac 
band-crossing point.

\end{abstract}


\maketitle

\section{Introduction}

Recent studies on fidelity and fidelity susceptibility\ct{zanardi06, venuti07, giorda07, gurev08} ($\chi_F$) near a quantum critical point have contributed to
 a deeper understanding  of a quantum phase transition\ct{sachdev99,chakrabarti96,continentino} from the viewpoint of quantum information theory. Fidelity is  the measure of overlap of two neighbouring ground 
states of a quantum Hamiltonian in the parameter space. Fidelity susceptibility provides quantitatively the rate of change of the ground state under an infinitesimal 
variation of the parameters of the Hamiltonian. Since the ground state of  a quantum many-body system exhibits different types of symmetries on either
 side of a quantum critical point\ct{sachdev99} (QCP), a sharp
drop of fidelity is observed right there. At the same time, the fidelity susceptibility usually diverges in a power law fashion with the system size where the exponent is given in terms of the
 quantum critical exponents\ct{venuti07, giorda07, capponi10, alet09, grandi10,gritsev09, lin09, gurev08, shigu08, zanardi, rams10, zhou}. In recent years a series of works have been directed to
 understanding the connection between fidelity susceptibility to quantum phase transition at critical\ct{venuti07, giorda07, capponi10, alet09, grandi10,gritsev09, lin09, gurev08, shigu08, zanardi, rams10, zhou} 
and multicritical points\ct{mukherjee10}. Studies of fidelity per site \ct{gurev08, zhou08}, reduced fidelity\ct{gurev08, prosen03, ma08, johannesson09} 
and geometric phase, which is also closely related to fidelity susceptibility \ct{venuti07}, near quantum critical \ct{pachos05, zhu06} and multicritical\ct{patra11} 
points have also been interesting areas of research.

In this paper, we extend the investigation on fidelity susceptibility to the case of an anisotropic quantum critical point (AQCP) and highlight the marked difference with the
corresponding  studies on an isotropic QCP. An interesting realization of an AQCP is seen in semi-Dirac band crossing points\ct{pardo09,singh09} where the 
energy gap scales linearly with momentum along one spatial direction but quadratically along others unlike Dirac points in
graphene where a gap opens linearly along both the directions\ct{castro09}. The possibility of such a semi-Dirac point has been reported recently\ct{pardo09,volovik01} using 
a three unit
 cell slab of $\rm {VO}_2$ confined within insulating $\rm {TiO}_2$ and also in liquid ${\rm He^3}$.
 A series of works on low energy properties of a system with a semi-Dirac point
has already been reported \ct{singh09, pardo10, delplace10}. It is to be noted that the scaling of defect density following a slow quench across a QCP, namely the 
Kibble Zurek scaling \ct{zurek05, polkovnikov05, dziarmaga05, levitov06, mukherjee07, sen08, dziarmaga_09, aeppli10, polkovnikov_rmp}, has also been generalized to an AQCP 
using a semi-Dirac Hamiltonian \ct{dutta10}. An AQCP can also be realized at the edge of the gapless region of a two dimensional Kitaev model in a 
honeycomb lattice\ct{kitaev03,mondal08} for which the Kibble Zurek scaling has also been proposed\ct{suzuki10}. 

Let us consider a  $d$-dimensional quantum mechanical Hamiltonian $H(\lambda)$ designated by a parameter $\lambda$. For two ground state
wavefunctions $\psi_0(\lambda)$ and $\psi_0 (\lambda + \delta \lambda)$ infinitesimally separated in the parameter space ($\delta \lambda \to 0$),
 we can define fidelity ($F$) as\ct{zanardi06, venuti07, giorda07, gurev08}
\ba
F = \left|\langle \psi_0(\lambda)|\psi_0 (\lambda + \delta \lambda) \rangle\right| \approx 1 - \frac{\delta \lambda^2}{2} \chi_F (\lambda) + \cdots
\label{fidelity}
\ea
where the fidelity susceptibility $\chi_{\rm F}$ is the first non-vanishing term in the expansion of fidelity. The scaling behavior of $\chi_{\rm F}$ at a QCP is well
 established \ct{venuti07,gurev08,shigu08}.  Let us choose the Hamiltonian to be of the form  $H = H_0 + \lambda H_I$. Here $H_0$ is the Hamiltonian describing a QCP 
at $\lambda = 0$ while $H_I\equiv \partial_\lambda H\bigr|_{\lambda=0}$ is the perturbation not commuting with $H_0$. 
One can relate the fidelity susceptibility density $\left(\chi_{\rm f} = 1/L^d \chi_{\rm F} \right)$ to the connected imaginary time ($\tau$) correlation function of the
 perturbation $H_I(\tau)$ using the relation \cite{venuti07} 
\be
\chi_{\rm f}(\lambda) = \frac{1}{L^d} \chi_{\rm F} = \frac{1}{L^d}\int_0^\infty \tau \langle H_I(\tau) H_I(0)\rangle_c d\tau.
\ee
Using  dimensional analysis in Eq.~(2), we get that the scaling dimension of $\chi_{\rm f}$ is given by ${\rm dim}[\chi_{\rm f}]=2\Delta_{H_I}-2 z + d$ where $z$
is the dynamical exponent associated with the QCP and $\Delta_{H_I}$ is the scaling dimension of the operator $H_I$.
Clearly a negative value of the scaling dimension leads to a  fidelity susceptibility diverging  with the system size $L$ at the QCP  as
$\chi_{\rm f} (\lambda=0) \sim L^{2z - d - 2\Delta_{H_I}}$.  A positive value, on the other hand, implies a singular $\chi_{\rm f}$ though
the singular behavior appears as a subleading correction to  a nonuniversal constant\ct{grandi10}. 
A marginal or relevant perturbation $H_I$ (so that $\lambda H_I$ scales as the energy) allows us to make an additional simplification
 coming from $\Delta_{H_I} = z-1/\nu$ so that at the critical point~\ct{alet09, gritsev09, capponi10, grandi10}
\be
\chi_{\rm f}\sim L^{2/\nu-d}.
\label{susc_1}
\ee
Further we get a cross-over from system size dependence to $\lambda$ dependence when the correlation length $\xi \sim \lambda^{-\nu}$ becomes of the order of
 system size:
\be
\chi_{\rm f}\sim |\lambda| ^{\nu d -2}.
\label{susc_2}
\ee 
These asymptotics are dominant for $d\nu<2$ and subleading for $d\nu>2$, while at $d\nu=2$ there are additional logarithmic singularities~\ct{gritsev09,grandi10}.

In the following analysis, we show that the general scaling of
 fidelity susceptibility valid near an isotropic QCP gets modified due to the anisotropy in critical behavior.  The changed scaling 
form  naturally includes  the  correlation length exponents along the different spatial directions, namely $\nu_{||}$ and $\nu_{\bot}$. In addition, for a finite 
system with linear dimension $L_{||}$ ($L_{\bot}$) in the parallel (perpendicular) directions, the maximum value of $\chi_{\rm f}$ increases 
with $L_{||}$ in the limit of small $L_{||}$ ($L_{||}^{1/\nu_{||}} \ll L_{\bot}^{1/\nu_{\bot}}$) only. In contrast, for higher values of 
$L_{||}$ ($L_{||}^{1/\nu_{||}} \gtrsim L_{\bot}^{1/\nu_{\bot}}$), we observe a crossover and $\chi_{\rm f}$ scales with $L_{\bot}$.
We also study the defect density, and heat density following a rapid quantum quench starting from an anisotropic quantum critical point and relate them through a 
generalized fidelity susceptibility \ct{grandi10}. We also highlight the connection to the Kibble-Zurek Scaling for the defect density following a slow
quench through an AQCP and retrieve the scaling relations derived previously\ct{dutta10, suzuki10}.

The paper is organized as follows: section II provides a general scaling relation of $\chi_{\rm f}$ associated with an AQCP.  We do also propose the 
same for the heat density and the defect density following a general quench starting from the AQCP, and relate them through a generalized fidelity susceptibility density. In section III we have taken a model
 Hamiltonian which shows an AQCP occurring in the physical systems described above and confirm our scaling predictions using exact analytical and numerical methods.
 Concluding remarks are presented in section IV.

\section{General Scaling Relations}
\subsection{Fidelity Susceptibility}

Let us consider a $d$ dimensional quantum Hamiltonian showing an AQCP at $\lambda = 0$. The correlations length exponent is $\nu = \nu_{||}$ along $m$ spatial directions and 
$\nu = \nu_{\bot}$ along rest of the ($d - m$) directions, called the parallel and perpendicular directions, respectively. The fidelity susceptibility as obtained from 
adiabatic perturbation theory\ct{polkovnikov05,ortiz08} is of  the form \ct{gurev08}
\ba
\chi_{\rm F} = \sum_{n \ne 0}\frac{|\langle \psi_n | \frac{\partial H}{\partial \lambda} |\psi_0 \rangle|^2}{(E_n - E_0)^2},
\ea
so that  for a finite system with linear dimension
 $L_{||}$ ($L_{\bot}$) in the parallel (perpendicular) directions, the corresponding fidelity susceptibility density ($\chi_{\rm f}$) can be written as
\be
\chi_{\rm f} = \frac{1}{L_{||}^m L_{\bot}^{d - m}} \chi_{\rm F} = \frac{1}{L_{||}^m L_{\bot}^{d - m}} 
\sum_{n \ne 0}\frac{|\langle \psi_n | \frac{\partial H}{\partial \lambda} |\psi_0 \rangle|^2}{(E_n - E_0)^2}.
\label{chif_1}
\ee
Here, $E_0$ and $E_n$ denote the energy of the ground state and $n$th energy level, respectively.
We can contrast the above equation (\ref{chif_1}) with the specific heat density (the second derivative of the ground state energy density 
($E_0/L^m_{||}L^{d - m}_{\bot}$)) given by \ct{capponi10}
\ba
\chi_E &=& -\frac{1}{L_{||}^m L_{\bot}^{d-m}}\partial^2 E_0/\partial \lambda^2 \non \\
&\sim& \frac{1}{L_{||}^m L_{\bot}^{d - m}} \sum_{n \ne 0}\frac{|\langle \psi_n | \frac{\partial H}{\partial \lambda} |\psi_0 \rangle|^2}{E_n - E_0}.
\label{chiE_1}
\ea
Comparison of Eqs.~(\ref{chif_1}) and (\ref{chiE_1}) indicates that near an AQCP, a much stronger divergence of $\chi_{\rm f}$  as compared to $\chi_{\rm E}$ is expected; this is
due to the higher power of energy difference  term in the denominator  of  $\chi_{\rm f}$. 

We note that near  an AQCP, the specific heat $\chi_{E} \sim |\lambda|^{-\alpha}$ where below the upper critical dimension the exponent $\alpha$ satisfies  a modified
hyperscaling relation\ct{continentino,binder89, bhattacharjee97}  $2 - \alpha = \nu_{||} m + \nu_{\bot} (d - m) + \nu_{||} z_{||}$. In the limit of large 
$|\lambda|$ ($|\lambda| \gg L_{||}^{-1/\nu_{||}}, L_{\bot}^{-1/\nu_{\bot}}$), $\chi_E$ scales as
\ba
\chi_E \sim |\lambda|^{-\alpha} \sim |\lambda|^{\nu_{||} m + \nu_{\bot} (d - m) + \nu_{||} z_{||} - 2 }.
\label{chiE_2}
\ea
Now, in the same limit the scaling of the fidelity susceptibility density is given by\ct{capponi10}
\ba
\chi_{\rm f} &=& \frac{1}{L_{||}^m L_{\bot}^{d - m}} \sum_{n \ne 0}\frac{|\langle \psi_n | \frac{\partial H}{\partial \lambda} |\psi_0 \rangle|^2}{(E_n - E_0)^2}
 \sim \frac{\chi_E}{|E_n - E_0|}\non \\ &\sim& |\lambda|^{\nu_{||} m + \nu_{\bot} (d - m) - 2} ~~~ \text{($|\lambda| \gg L_{||}^{-1/\nu_{||}}, L_{\bot}^{-1/\nu_{\bot}}$)}.
\label{chif_2}
\ea
In deriving Eq.~(\ref{chif_2})  we have used Eq.~(\ref{chiE_2}) and the fact that near the AQCP, $E_n - E_0 \sim |\lambda|^{\nu_{||} z_{||}} = |\lambda|^{\nu_{\bot} z_{\bot}}$. In the special case 
of $\nu_{||} = \nu_{\bot} = \nu$, we retrieve the expected scaling relation  $\chi_{\rm f} \sim \lambda^{\nu d - 2}$ valid near an isotropic quantum critical 
point\ct{grandi10, gritsev09} (see eq. (4)). 

On the other hand, right at the AQCP ($\lambda = 0$), and in the limit $L_{||}^{1/\nu_{||}} \ll L_{\bot}^{1/\nu_{\bot}}$, 
$\chi_{\rm f}(\lambda = 0)$ scales with the system size $L_{||}$ as
\ba
\chi_{\rm f} (\lambda = 0) \sim L_{||}^{\frac{2}{\nu_{||}} - \frac{\nu_{\bot}}{\nu_{||}} (d - m) - m} ~~~\text{$(L_{||}^{1/\nu_{||}} \ll L_{\bot}^{1/\nu_{\bot}})$}.
\label{scale1}
\ea
However, in the opposite limit $L_{||}^{1/\nu_{||}} \gg L_{\bot}^{1/\nu_{\bot}}$, $\chi_{\rm f}(\lambda = 0)$ instead starts scaling with 
$L_{\bot}$, and Eq.~(\ref{scale1}) gets modified to
\ba
\chi_{\rm f} (\lambda = 0) \sim L_{\bot}^{\frac{2}{\nu_{\bot}} - \frac{\nu_{||}}{\nu_{\bot}}m - (d - m)}~~~\text{$(L_{||}^{1/\nu_{||}} \gg L_{\bot}^{1/\nu_{\bot}})$}.
\label{scale2}
\ea
Clearly, the special condition $L_{||}^{1/\nu_{||}} \sim L_{\bot}^{1/\nu_{\bot}}$ yields
\ba
\chi_{\rm f} (\lambda = 0) \sim L_{||}^{\frac{2}{\nu_{||}} - \frac{\nu_{\bot}}{\nu_{||}} (d - m) - m} \sim L_{\bot}^{\frac{2}{\nu_{\bot}} - \frac{\nu_{||}}{\nu_{\bot}}m - (d - m)} .
\label{scale3}
\ea
The above scalings in Eqs.~(\ref{scale1} - \ref{scale3}) suggest $\chi_{\rm f} (\lambda = 0)$ initially increases with $L_{||}$ until $L_{||}^{1/\nu_{||}} \sim L_{\bot}^{1/\nu_{\bot}}$. 
Beyond which $\chi_{\rm f}(\lambda = 0)$ becomes independent of $L_{||}$ and saturates to a constant value. However, in this limit the fidelity susceptibility density 
scales with $L_{\bot}$, as shown in Eq.~(\ref{scale2}).

An alternative way of arriving at the above scalings is by the use of correlation functions\ct{venuti07,giorda07, lin09, gurev08, gritsev09,capponi10, alet09}:
\ba
\chi_{\rm f} = \frac{1}{L_{||}^m L_{\bot}^{d-m}} \int^{\infty}_0 {\tau} \langle H_I(\tau) H_I(0)\rangle_c d\tau,
\label{corr}
\ea
where we define $$H_I(\tau) = e^{H \tau} H_I e^{-H \tau}$$ and
 $$\langle H_I(\tau) H_I(0)\rangle_c = \langle H_I(\tau) H_I(0)\rangle - \langle H_I(\tau)\rangle \langle H_I(0)\rangle,$$ with $\tau$ being the imaginary time.
 For a relevant perturbation $\lambda H_I$ should scale as the energy, so that $H_I \sim \lambda^{\nu_{||} z_{||} - 1}$. Using the relation 
$\tau \sim L_{||}^{z_{||}}$ and $L_{||,\bot} \sim \lambda^{-\nu_{||,\bot}}$, we get the scaling of $\chi_{\rm f}$ from Eq.~(\ref{corr})
given by
\ba
\chi_{\rm f} \sim |\lambda|^{\nu_{||} m + \nu_{\bot} (d - m) - 2},
\ea
which is identical to Eq.~(\ref{chif_2}).

\subsection{Heat and defect density following a sudden quench}
In this section we study a sudden quench\ct{santoro09,santoro10} of a quantum system of amplitude $\lambda$, starting from the AQCP. The quantities of interest are 
defect density\ct{dziarmaga_09,aeppli10, polkovnikov_rmp} ($n_{ex}$) and heat density\ct{grandi10} ($Q$) generated in the process. Advantage of using heat density, 
or the excess energy above the new ground state, is that it can be defined even for non-integrable systems. On the other hand, for an integrable system with non-interacting quasi-particles, it is 
useful to define defect density, which is a measure of the density of excited quasi-particles generated in the system.

As $\lambda$ is suddenly increased from $\lambda = 0$ to its final value $\lambda$, all the momentum modes 
$k_{||} \lesssim \lambda^{\nu_{||}}$ and $k_{\bot} \lesssim \lambda^{\nu_{\bot}}$ get excited with excitation energy 
$\sim \lambda^{\nu_{||} z_{||}} = \lambda^{\nu_{\bot} z_{\bot}}$ for each mode. 
This gives an excitation energy density or heat density of the form
\ba
Q \sim \lambda^{\nu_{||} m + \nu_{\bot} (d - m) + \nu_{||} z_{||}}.
\label{eq_heat}
\ea

Defect density is related to the probability of excitation,
 which in turn can be expressed in terms of fidelity susceptibility\ct{gu09,grandi10, gritsev09}. Following the above argument one finds that
\ba
n_{ex} \sim \lambda^2 \chi_{\rm f} \sim \lambda^{\nu_{||} m + \nu_{\bot} (d - m)}.
\label{eq_kink}
\ea
Eq.~(\ref{eq_kink}) can also be derived by noticing that for a sudden quench of amplitude $\lambda$, all the momentum modes $k_1 \lesssim \lambda^{\nu_{||}}$
and $k_2 \lesssim \lambda^{\nu_{\bot}}$ get excited with unit probability, giving $n_{ex} \sim \lambda^{\nu_{||} m + \nu_{\bot} (d - m)}$. 

\subsection{Generalized fidelity susceptibility density}
In this section we deal with a generic quench from an AQCP at time $t=0$ given by

\be
\lambda(t) = \delta\frac{t^r}{r!} \Theta (t),
\label{gen_quench}
\ee
where $\delta$ is a small parameter, and $\Theta$ is
the step function\ct{grandi10}. The case $r = 0$ denotes a rapid quench of amplitude $\delta$; the case $r = 1$ implies a slow linear quench with a rate $\delta$ and so on. 
In all these cases the limit  $\delta\to 0$ is considered to signify a slow adiabatic time evolution. If the system is initially in the ground
 state, the transition probability to the instantaneous excited state as obtained from the adiabatic perturbation theory is given by 
\ba
P_{ex} &=& \delta^2 \sum_{n \ne 0}\frac{|\langle \psi_n | \frac{\partial H}{\partial \lambda} |\psi_0 \rangle|^2}{(E_n - E_0)^{2r + 2}} \non\\
&=& \delta^2 L_{||}^m L_{\bot}^{d - m} \chi_{\rm 2r + 2},
\ea
which leads to a density of defect of the form
\ba
n_{ex} = \frac{1}{ L_{||}^m L_{\bot}^{d - m}} P_{ex} = \delta^2  \chi_{\rm 2r + 2}.
\ea
In the above, we have used the definition of a generalized fidelity susceptibility density $\chi_{\rm l}$ given by\ct{grandi10} 
\ba
\chi_{\rm l} = \frac{1}{L_{||}^{m}}\frac{1}{L_{\bot}^{d-m}} \sum_{n \ne 0}\frac{|\langle \psi_n | \frac{\partial H}{\partial \lambda} |\psi_0 \rangle|^2}{(E_n - E_0)^l}.
\label{gen_suscp}
\ea
From Eq.~(\ref{gen_suscp}), one finds that
 $\chi_1$ stands for the specific density $\chi_{\rm E}$ while  $\chi_2$ is the fidelity susceptibility density $\chi_{\rm f}$; $\chi_4$, on the other hand, yields
the excitation probability following a
 slow linear quench starting from an AQCP.

In the same spirit as in Eq.~(2),  a general $\chi_{\rm l}$ can also be expressed in terms of time dependent connected 
correlation functions given by
\ba
\chi_{\rm l} = \frac{1}{L_{||}^m L_{\bot}^{d - m} (l - 1)!} \int^\infty_0 {\tau^{l - 1} \langle H_I(\tau) H_I(0) \rangle_c d\tau}.
\label{chil}
\ea
Now, using $\lambda \sim L_{||}^{-1/\nu_{||}} \sim L_{\bot}^{-1/\nu_{\bot}}$ and $t \sim L_{||}^{z_{||}} \sim L_{\bot}^{z_{\bot}}$ in Eq.~(\ref{gen_quench})
leads to the scaling relations $L_{||} \sim \delta^{-\frac{\nu_{||}}{1 + \nu_{||} z_{||} r}}$,
 $L_{\bot} \sim \delta^{-\frac{\nu_{\bot}}{1 + \nu_{\bot} z_{\bot} r}} = \delta^{-\frac{\nu_{\bot}}{1 + \nu_{||} z_{||} r}}$. These suggest that one can further
conclude  $H_I \sim \lambda^{\nu_{||} z_{||} - 1} 
\sim \delta^{\frac{\nu_{||} z_{||} - 1}{1 + r \nu_{||} z_{||}}}$,
and $\tau \sim L_{||}^{z_{||}} \sim \delta^{-\frac{\nu_{||} z_{||}}{1 + \nu_{||} z_{||} r}}$. Substituting for $L_{||}$, $L_{\bot}$, $H_I$ and $\tau$ in Eq.~(\ref{chil}) 
with $l = 2r + 2$  one gets in  the limit $\delta \gg L_{||}^{-\frac{1}{\nu_{||}} - z_{||} r}, L_{\bot}^{-\frac{1}{\nu_{\bot}} - z_{\bot} r}$
\ba
\chi_{\rm 2r + 2} &\sim& \delta^{\frac{\nu_{||} m + \nu_{\bot} (d - m) - 2 - 2 \nu_{||} z_{||} r}{1 + \nu_{||} z_{||} r}} .
\label{chil_scale}
\ea
Therefore for a generic quench from an AQCP scaling of  defect density gets modified to
\ba
n_{ex} \sim \delta^{\frac{\nu_{||} m + \nu_{\bot} (d - m)}{\nu_{||} z_{||} r + 1}} ~~~
\text{$\left(\delta \gg L_{||}^{-\frac{1}{\nu_{||}} - z_{||} r}, L_{\bot}^{-\frac{1}{\nu_{\bot}} - z_{\bot} r}\right)$},
\label{slow_defect}
\ea
 while the  corresponding heat density scales as
\ba
Q \sim \lambda^{\nu_{||} z_{||}} n_{ex}\sim \delta^{\frac{\nu_{||} m + \nu_{\bot} (d - m) + \nu_{||} z_{||}}{\nu_{||} z_{||} r + 1}}. 
\label{slow_heat}
\ea
The expressions for $n_{ex}$ and $Q$ match exactly with the same for fast quench (Eqs. (\ref{eq_heat},\ref{eq_kink})) if we put $r = 0$ and $\delta = \lambda$, whereas, the 
case $r = 1$ correctly reproduces the values for a slow linear quench starting from the AQCP\ct{dutta10,suzuki10}.

The scaling relations presented above are valid as long as the corresponding exponents do not exceed two. Otherwise contributions  from 
short wavelength modes become dominant and hence the low energy singularities associated with the critical point become subleading\ct{gritsev09, grandi10}.

\section{Model and Hamiltonian}
\begin{figure}[htb]
\includegraphics[height=2.0in,width=3.1in, angle = 0]{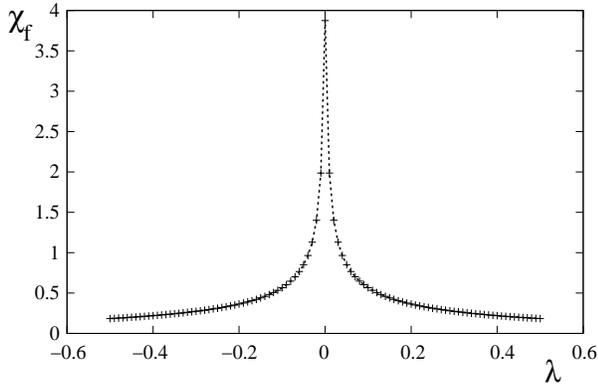}
\caption{Variation of $\chi_{\rm f}$ with $\lambda$, as obtained numerically for $L_{||} = 10000$, $L_{\bot} = 1000$, $\nu_{||} = 1/2$, $\nu_{\bot} = 1$, $d = 2$ and $m = 1$.
 $\chi_{\rm f}$ peaks at the AQCP, and falls as $|\lambda|^{-1/2}$, as predicted in Eq.~(\ref{chif_2}).}
\label{chi_lambda}
\end{figure}
\begin{figure}[htb]
\includegraphics[height=2.0in,width=3.1in, angle = 0]{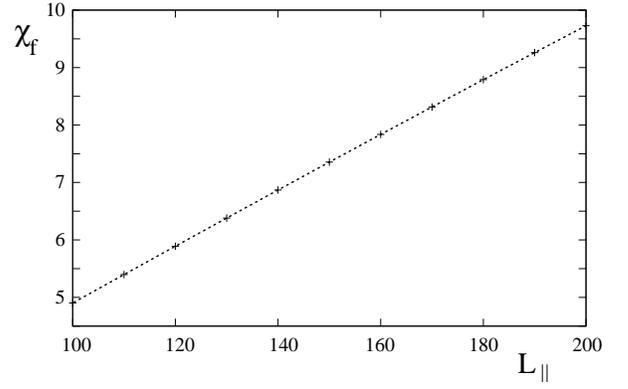}
\caption{Variation of $\chi_{\rm f} (\lambda = 0)$ with $L_{||}$ as obtained numerically for $L_{\bot} = 100000$, $\nu_{||} = 1/2$, $\nu_{\bot} = 1$, $d = 2$ and $m = 1$. $\chi_{\rm f}$ diverges as
$\chi_{\rm f} \sim L_{||}$, in agreement with the scaling given in Eq.~(\ref{scale1}).}
\label{fig_Lparal}
\end{figure}
\begin{figure}[htb]
\includegraphics[height=2.0in,width=3.1in, angle = 0]{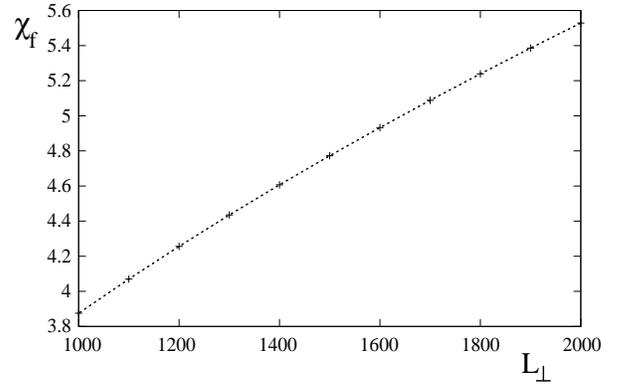}
\caption{Variation of $\chi_{\rm f} (\lambda = 0)$ with $L_{\bot}$ as obtained numerically for $L_{||} = 10000$, $\nu_{||} = 1/2$, $\nu_{\bot} = 1$, $d = 2$ and $m = 1$. $\chi_{\rm f}$ diverges as
$\chi_{\rm f} \sim L_{\bot}^{1/2}$ as expected from the scaling Eq.~(\ref{scale2}).}
\label{fig_Lperp}
\end{figure}
\begin{figure}[htb]
\includegraphics[height=2.0in,width=3.1in, angle = 0]{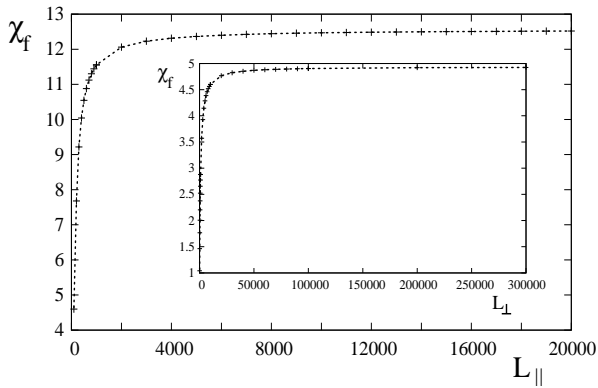}
\caption{Variation of $\chi_{\rm f} (\lambda = 0)$ with $L_{||}$ as obtained numerically for $\nu_{||} = 1/2$, $\nu_{\bot} = 1$, $d = 2$, $m = 1$ and $L_{\bot} = 10000$. 
$\chi_{\rm f}$ saturates at $L_{||}^2 \gtrsim L_{\bot}$, as expected from Eqs.~(\ref{chi_0}, \ref{eq_susc}). Inset shows Variation of $\chi_{\rm f}$ with $L_{\bot}$ when $L_{||}$ kept fixed at 
$L_{||} = 100$. $\chi_{\rm f}$ saturates at $ L_{\bot} \gtrsim L_{||}^2$.}
\label{sat}
\end{figure}
\begin{figure}[htb]
\includegraphics[height=2.0in,width=3.1in, angle = 0]{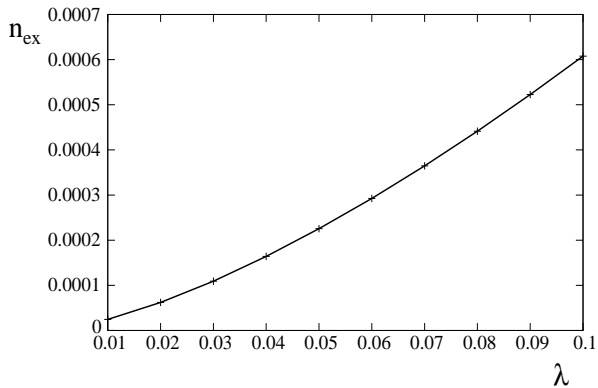}
\caption{Kink density $n_{ex}$ as a function of $\lambda$ as obtained numerically for $L_{||} = L_{\bot} = 1000$, $\nu_{||} = 1/2$, $\nu_{\bot} = 1$, $d = 2$, $m = 1$ 
and $\lambda \gg L^{-1/\nu_{||}}_1, L_{\bot}^{-1/\nu_{\bot}}$. $n_{ex}$ varies as $n_{ex} \sim \lambda^{3/2}$, as predicted in Eq.~(\ref{eq_kink}).}
\label{fig_kink}
\end{figure}
We illustrate the above analytical predictions using the representative case of a semi-Dirac point in spatial dimension $d = 2$. In this case, $\nu_{||} = 1/2$,
 $\nu_{\bot} = 1$ and $d =2, m = 1$. In the momentum $k$ space the Hamiltonian near a semi-Dirac point can be written as the direct product of $2 \times 2$ 
Hamiltonians given by \ct{singh09, dutta10, suzuki10}
\ba
H_k &=& \left[ \begin{array}{cc} \lambda & k_{||}^{2} + ik_{\bot} \\
k_{||}^2 - ik_{\bot} & -\lambda \end{array} \right] .
\label{Hamil}
\ea
The fidelity susceptibility density near the semi-Dirac point ($\lambda = 0$) can be written as as
\ba
\chi_{\rm f} &=& \frac{1}{\pi^2} \int_{\pi/L_{||}}^{\pi} \int_{\pi/L_{\bot}}^{\pi} {\frac{k_{||}^4 + k_{\bot}^2}{(\lambda^2 + k_{||}^4 + 
k_{\bot}^2)^2}} dk_{||} dk_{\bot}.
\ea
Rescaling $k_{||}/\sqrt{\lambda} = x_1$, $k_{\bot}/\lambda = x_2$ and taking the limit $\lambda \gg L_{||}^{-2}, L_{\bot}^{-1}$ we get
\ba
\chi_{\rm f} &=& \frac{1}{|\lambda|^{1/2}\pi^2} \int_{\pi/\sqrt{\lambda}L_{||}}^{\pi/\sqrt{\lambda}} 
\int_{\pi/\lambda L_{\bot}}^{\pi/\lambda} {\frac{x_1^4 + x^2_2}{\left(1 + x_1^4 + x^2_2\right)^2}} dx_1 dx_2 \non \\
&\approx& \frac{1}{|\lambda|^{1/2}\pi^2} \int_{0}^{\infty} \int_{0}^{\infty} {\frac{x_1^4 + x^2_2}{\left(1 + x_1^4 + x^2_2\right)^2}} dx_1 dx_2 \non \\
&\sim& |\lambda|^{-1/2},
\label{general_chi}
\ea
which shows that divergence of $\chi_{\rm f}$ at $\lambda \to 0$ (see Fig. (\ref{chi_lambda})) and exponent (1/2), and are in complete agreement with Eq.~(\ref{chif_2}) 
for $d = 2$, $ m = 1$, $\nu_{||} = 1/2$ and $\nu_{\bot} = 1$.

\begin{figure}[htb]
\includegraphics[height=2.0in,width=3.1in, angle = 0]{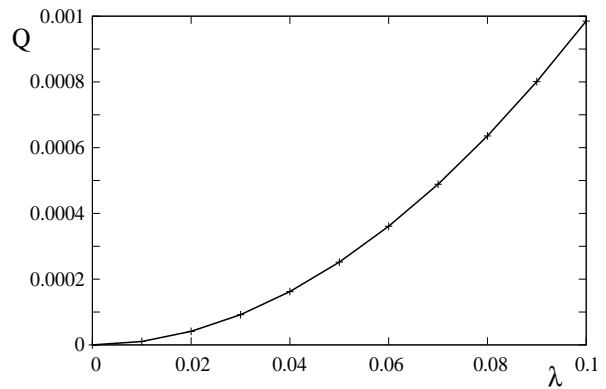}
\caption{Heat density $Q$ as a function of $\lambda$ as obtained numerically for $L_{||} = L_{\bot} = 1000$, $\nu_{||} = 1/2$, $\nu_{\bot} = 1$, $d = 2$, $m = 1$ 
and $\lambda \gg L^{-1/\nu_{||}}_1, L_{\bot}^{-1/\nu_{\bot}}$. $Q$ follows the perturbative scaling law $Q \sim \lambda^2$, as discussed in the text.}
\label{fig_heat}
\end{figure}

Right at the AQCP ($\lambda = 0$), we have. 
\ba
\chi_{\rm f}(\lambda = 0) &\approx& \frac{1}{\pi^2} \int_{\pi/L_{||}}^{\infty} \int_{\pi/L_{\bot}}^{\infty} {\frac{1}{k_{||}^4 + k_{\bot}^2}} dk_{||} dk_{\bot}.
\ea
The scalings $k_{\bot} = p k_{||}^2$ or 
$k_{||} = q \sqrt{k_{\bot}}$ simplify the above integral to
\ba
\chi_{\rm f} &=& \frac{1}{\pi^2}\int^{\infty}_{\pi/L_{||}} \frac{d k_{||}}{k_{||}^2} \int^{\infty}_{\pi/L_{\bot} k_{||}^2} \frac{dp}{p^2 + 1} \non \\
&=& \frac{1}{\pi^2}\int^{\infty}_{\pi/L_{\bot}} \frac{d k_{\bot}}{k_{\bot}^{3/2}} \int^{\infty}_{\pi/L_{||}\sqrt{k_{\bot}}} \frac{dq}{q^4 + 1}.
\label{chi_0}
\ea
In the limit $L_{||}^2 \ll L_{\bot}$ (or $L_{||}^{2} \gg L_{\bot}$) we can approximate $\pi/L_{\bot}k^2_{||}$ (or $\pi/L_{||} \sqrt{k_{\bot}}$) to zero, 
so that the scalings in Eq.~(\ref{chi_0}) depends on one of the length scales, and we get
\ba
\chi_{\rm f} (\lambda = 0) &\sim& L_{||} ~~~~~~~ \text{(for $L_{||}^{2} \ll L_{\bot}$)}, \non \\
&\sim& L_{\bot}^{1/2} ~~~~~ \text{(for $L_{||}^{2} \gg L_{\bot}$)}.
\label{eq_susc}
\ea
Numerical verifications for the scalings of $\chi_{\rm f}$ with $L_{||}$ and $L_{\bot}$ discussed in Eq.~(\ref{eq_susc}) are provided in 
figures (\ref{fig_Lparal} - \ref{sat}). Extending our analysis of $\chi_{\rm f}$ to find the defect density following a fast quench starting from the AQCP ($\lambda = 0$), we arrive at the scaling
$n_{ex} \sim \lambda^2 \chi_{\rm f} \sim \lambda^{3/2}$. This relation is in perfect agreement with Eq.~(\ref{eq_kink}) and is verified numerically as shown in 
Fig. (\ref{fig_kink}). 
However, scaling analysis of heat density Eq.~(\ref{eq_heat}) predicts $Q \sim |\lambda|^{2.5}$, which is subleading to the quadratic form $Q \sim \lambda^{2}$ 
arising from contributions of short wavelength modes. This leads to the scaling relation $Q \sim \lambda^2$. This quadratic scaling is also checked numerically 
in Fig.~(\ref{fig_heat}).  

\section{Conclusions}

We have studied the scaling behavior of fidelity susceptibility near an anisotropic quantum critical point. 
Anisotropic critical behaviour modifies the general scaling form of $\chi_{\rm f}$. In particular,
 both $\nu_{||}$ and $\nu_{\bot}$ appear in the scaling. In addition, even though the maximum value of $\chi_{\rm f}$ scales with $L_{||}$ in the limit of small 
$L_{||}$, at higher values of the same  a cross-over is observed and $\chi_{\rm f}$ starts scaling with $L_{\bot}$ instead. We also propose the scaling relations for the defect 
density and heat density  following a generic quantum quench starting from an AQCP and relate them through a generalized fidelity susceptibility. We have verified our general 
scaling predictions both numerically and analytically using the illustrative example of a Hamiltonian showing a semi-Dirac point. Interestingly, we show that the
heat density following a rapid quench starting from a two-dimensional semi-Dirac point varies quadratically with the amplitude and the scaling arising due to low-energy
critical modes appear only as a sub-leading correction.

\acknowledgements
The authors acknowledge U. Divakaran, A. Polkovnikov and R. R. P. Singh for collaboration in related works. A.D. acknowledges CSIR, New Delhi, for partial financial support.

\end{document}